\documentclass{optica-article}
\usepackage{graphicx}
\usepackage{amsmath}
\usepackage{orcidlink} 
\usepackage{tabularx}
\usepackage{subcaption}
\usepackage{dsfont}
\usepackage{float}   
\journal{opticajournal} 
\articletype{Research Article}
\usepackage{lineno}
\usepackage{soul}
\usepackage{fancyhdr}
\begin{document}
\title{Robustness of Quantum Correlations and Controlled Teleportation Fidelity under Noisy Channels}

\author{S. Archana \authormark{1}, Aditi Das \authormark{2}, Venkat Abhignan\,\orcidlink{0000-0002-6092-3670} \authormark{2,3,*}, and R. Muthuganesan\,\orcidlink{0000-0002-9188-1151}\authormark{1,*}}

\address{\authormark{1}Department of Physics and Nanotechnology, SRM Institute of Science and Technology, Kattankulathur - 603203, Tamil Nadu, India\\
\authormark{2}Qdit Labs Pvt. Ltd., Bengaluru 560092, Karnataka, India
\authormark{3}Department of Applied Physics and Materials Engineering, National Institute of Technology Patna, Bihar - 800005, India}

\email{\authormark{*}venkat.ph@nitp.ac.in, \authormark{*}rajendramuthu@gmail.com} 


\begin{abstract}
In distributed quantum computing, quantum networks provide a potential infrastructure for long-range quantum information processing. Controlled quantum teleportation (CQT) is a key practically relevant protocol that can connect two parties on a network and enables a third party, the controller to allow or abort the secure teleportation of an arbitrary single-qubit state between the two parties. We investigate the robustness of a three-qubit GHZ resource under bit-flip, phase-flip, depolarizing, and amplitude-damping noise. Using the Kraus-operator formalism, we characterize the noisy resource through multipartite quantum discord, genuine multipartite concurrence (GMC), localizable entanglement (LE), and CQT fidelity. We also outline a proof-of-concept photonic implementation for preparing the three-qubit GHZ resource and realizing the CQT protocol, which could enable experimental investigation of the effects of noise. Our results show that bit-flip and amplitude-damping noise generally preserve useful mutipartite quantum correlations over broader noise parameter ranges. Interestingly, multipartite discord exhibits freezing under bit-flip noise over a finite parameter regime. Moreover, discord can persist beyond the sudden death of genuine multipartite entanglement, revealing the distinct robustness of nonclassical correlations. We further demonstrate that CQT fidelity which characterizes the performance of CQT can exceed the classical threshold even after GMC vanishes, while LE more closely reflects the operational usefulness of the noisy resource. These findings highlight the complementary roles of quantum correlations and entanglement in assessing the resilience of GHZ-based protocols
\end{abstract}
\section{Introduction}
Quantum teleportation has emerged as a fundamental and highly useful protocol for quantum communication, enabling the transfer of quantum information between distant parties through pre-shared entanglement and classical communication \cite{PhysRevLett.70.1895,Pirandola2015}. Quantum networks extend this capability to distributed quantum systems, where entanglement distribution \cite{PhysRevLett.117.240503} and teleportation \cite{PhysRevLett.84.3482,Yonezawa2004,Yin2012,Ma2012} provide essential mechanisms for connecting remote quantum processors for supporting quantum communication and computation \cite{Kimble2008,8910635}. The concept of quantum teleportation was first introduced by Bennett \textit{et al.} in 1993, in which an unknown single-qubit state is transferred using an entangled two-qubit channel and classical communication \cite{PhysRevLett.70.1895}. Later, researchers began studying multipartite entangled states, such as three-qubit states, because they exhibit richer, more complex correlations \cite{RevModPhys.81.865}. Three-qubit and multipartite entangled states exhibit correlations that are not captured by pairwise entanglement alone, motivating their use in quantum communication and distributed information-processing protocols \cite{PhysRevA.61.052306}. Among multipartite entangled states, Greenberger–Horne–Zeilinger (GHZ) states stand out as particularly powerful resources \cite{Greenberger1989}, enabling CQT \cite{PhysRevA.58.4394}, quantum secret sharing \cite{PhysRevA.59.1829}, and multiparty quantum communication \cite{Wu2021,Li2023,Shu2025}. Compared with Bell-state-based schemes, the multipartite structure of GHZ states naturally supports communication protocols involving more than two parties, making it especially well-suited to network-based quantum communication architectures \cite{McCutcheon2016,Pant2019,PhysRevLett.124.060501,Pirandola2015}. In the standard GHZ-based CQT protocol considered here, a third party, Charlie, acts as a controller, and Bob requires Charlie’s measurement information, in addition to Alice’s measurement results, to reconstruct the teleported state \cite{Barasinski2018,PhysRevLett.122.170501,PhysRevA.99.012306}. The practical relevance of the CQT protocol is evident in experimental realizations using both continuous-variable \cite{Yonezawa2004} and discrete-variable GHZ states \cite{PhysRevLett.122.170501}. Additionally, CQT has been proposed using different physical platforms and multipartite entangled resources, including GHZ states generated in trapped-ion systems \cite{Xu2016}, Brown states implemented through cavity-QED architectures \cite{WANG2009}, quantum-dot systems \cite{PhysRevA.87.062337,Heo2017}, and GHZ-like entangled states \cite{Naseri2015}.

The performance of CQT is strongly influenced by the quality of the shared multipartite quantum resource between Alice, Bob, and Charlie. In realistic implementations, the entangled resource is inevitably exposed to environmental interactions during state preparation, transmission, and processing. Importantly, different qubits in the resource may experience distinct local environments, making the location of noise within the protocol a key factor in determining the protocol's robustness. Such interactions introduce decoherence and can transform an initially pure entangled state into a mixed state, thereby weakening the quantum correlations required for reliable teleportation \cite{RevModPhys.75.715,nielsen2000quantum}. Different noise channels affect quantum states in distinct ways: bit-flip noise alters the logical state of qubits; phase-flip noise destroys phase coherence of the state without energy loss; depolarising noise uniformly randomizes the quantum state ("white noise"); and amplitude-damping noise describes dissipative energy-loss processes. In photonic quantum communication, photon loss due to fiber attenuation is
closely related to amplitude-damping-type dynamics, as established through
a Kraus-operator description of single-photon propagation in lossy fibers
\cite{PhysRevA.105.062437}.  Since teleportation fidelity depends critically on preserving relevant quantum correlations, a thorough understanding of how these noise channels affect the protocol is essential for developing reliable quantum communication systems \cite{PhysRevA.92.012338}. Notably, while such noise is generally assumed to be detrimental \cite{PhysRevA.65.022302,PhysRevA.86.020304}, experimental \cite{PhysRevA.90.042332}, and theoretical studies \cite{PhysRevA.62.012311,PhysRevA.78.022334} have shown that interaction with a suitably engineered local environment can, counterintuitively, enhance teleportation fidelity rather than degrade it in the case of Bennett teleportation. 

 Furthermore, localizable entanglement, defined as the averaged conditional biseparable entanglement \cite{LocalizableEnt_Pop2005,EntCor_Verstraete2004}, provides a natural measure of the entanglement resource available for CQT \cite{Barasinski2018,PhysRevA.99.012306}. This is particularly relevant when the noisy GHZ-resource loses genuine multipartite entanglement, since CQT does not necessarily require a genuinely tripartite-entangled state. Previous theoretical work showed that mixed biseparable three-qubit states with nonzero localizable entanglement can still enable CQT with fidelity above the classical threshold \cite{Barasinski2018,7cfj-8klm}. This result was further supported experimentally through CQT from discrete-variable GHZ states using linear optical devices \cite{PhysRevLett.122.170501}. Motivated by these results, we use localizable entanglement, together with GME, to characterize how different noise channels affect the quantum resources underlying CQT.

Beyond entanglement-based measures, quantum systems can exhibit several distinct forms of nonclassical correlations, including nonlocality \cite{PhysicsPhysiqueFizika.1.195}, Einstein–Podolsky–Rosen (EPR) steering \cite{1935PhRv...47..777E,Schrödinger_1935}, and quantum discord \cite{PhysRevLett.88.017901,LHenderson_2001}, which characterize different aspects of quantumness. Nonlocality, EPR steering, and entanglement are commonly associated within a hierarchy of quantum correlations \cite{PhysRevLett.98.140402}, whereas quantum discord provides a complementary characterization of nonclassical correlations that can persist under noise even in separable states \cite{PhysRevA.80.024103}. This is particularly relevant for noisy multipartite resources, where environmental interactions can reduce genuine multipartite and localizable entanglement while other forms of nonclassical correlations can remain non-zero. Also, experiments have demonstrated that separable states with discord can outperform entangled states in specific tasks such as remote state preparation \cite{Dakić2012}. Examining discord alongside GME and localizable entanglement therefore provides a broader characterization of how noise modifies the quantum correlations of the GHZ resource which might be relevant to CQT. Discord is defined on the intuition that von Neumann measurements break the quantum correlations between the subsystems, and by comparing the mutual information of the original state with that of the resulting classically correlated state, one can quantify the total quantum correlations present with respect to the measured subsystem \cite{PhysRevA.84.042109,PhysRevA.82.034302}. For pure states, quantum discord coincides with the corresponding entanglement measure, whereas their relationship becomes more subtle for mixed states that may possess nonzero quantum discord despite having no entanglement \cite{RevModPhys.84.1655,Bera_2018}. Although genuine multipartite entanglement has been rigorously studied \cite{PhysRevA.83.062325,PhysRevA.86.062303,PhysRevLett.127.040403,PhysRevResearch.5.L032015} as an extension of original bipartite concurrence of Wootters \cite{PhysRevLett.78.5022,PhysRevLett.80.2245}, the characterization of multipartite discord based on its original definition \cite{PhysRevLett.88.017901,LHenderson_2001} has been quite recent \cite{PhysRevLett.124.110401,Gulati_2025}. 


The remainder of this paper is organized as follows. Sec. 2 outlines the experimentally relevant CQT protocol. The general analysis in CQT is presented initially and does not depend on a particular physical realization of the qubits. Further, its experimental relevance was illustrated using photonic implementations. In particular, a relevant polarization-based photonic realization of the three-qubit GHZ resource including a phase-stable cascaded-SPDC source for generating three-photon entanglement \cite{PhysRevA.105.063705} and linear-optical implementation of CQT protocol \cite{PhysRevLett.122.170501} were described. Sec. 3 describes the four noise channel models considered in this study and Sec. 4 presents the descriptions for the multipartite quantum discord, genuine multipartite entanglement, localizable entanglement and fidelity of CQT. Sec. 5 provides a detailed discussion of the results, including comparative analysis of all the quantum measures across different noise types. Sec. 6 summarizes the major conclusions and outlines directions for future work.

\section{Controlled Quantum Teleportation}

\begin{figure}[t]
    \includegraphics[width=\textwidth]{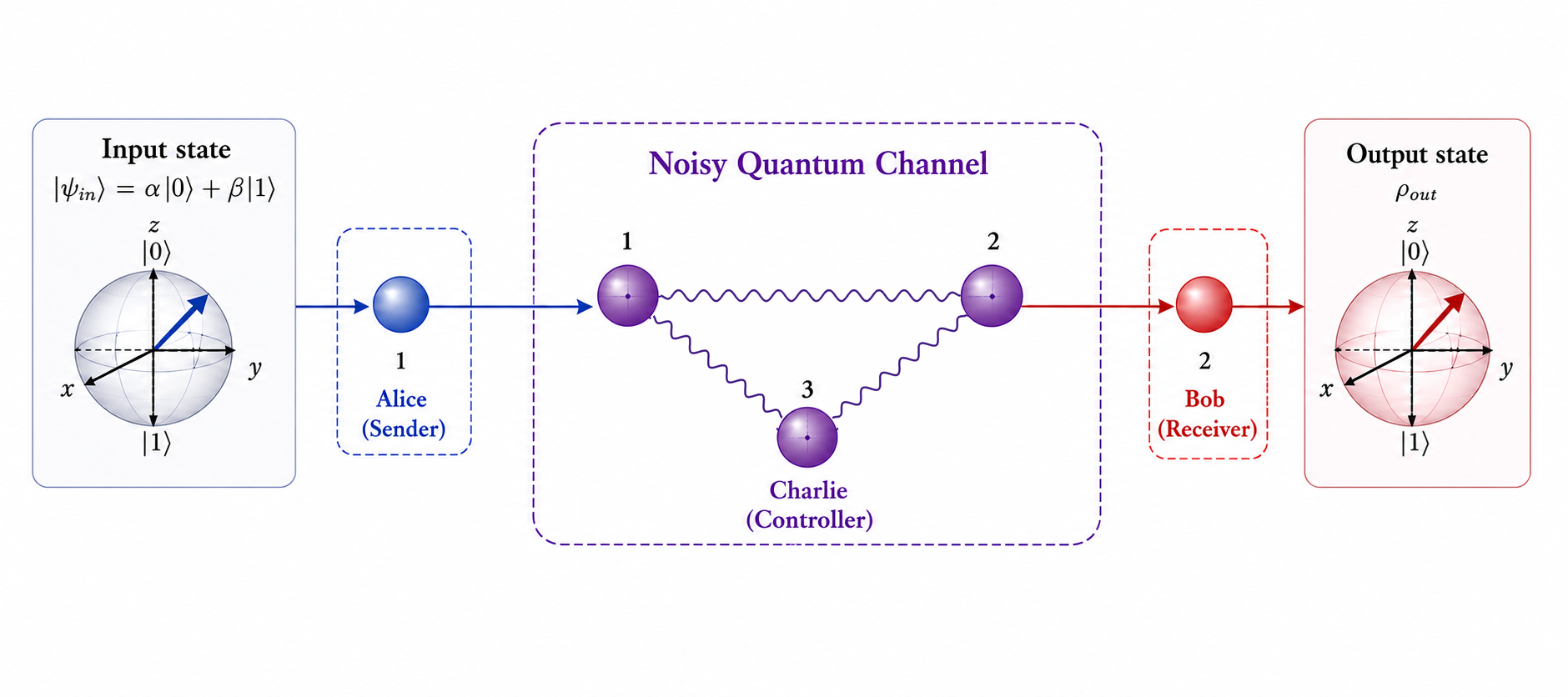}
    \caption{Schematic representation of controlled quantum teleportation of an arbitrary single-qubit state $|\psi_{\mathrm{in}}\rangle$. Alice, Bob, and Charlie hold qubits $1$, $2$, and $3$ of the shared resource $\rho_{GHZ}$, respectively using a noisy three-qubit quantum channel. Alice performs the sender-side Bell measurement between the input state and qubit $1$, Charlie performs the controller measurement on qubit $3$, and Bob applies the corresponding conditional operation to qubit $2$ to obtain the output state $\rho_{\mathrm{out}}$. The shared three-qubit channel is subject to bit-flip (BF), phase-flip (PF), depolarising (DP), and amplitude-damping (AD) noise.}
    \label{fig:noisy_CQT_Xstate}
\end{figure}

We describe the CQT protocol based on Refs. \cite{Barasinski2018,PhysRevLett.122.170501,PhysRevA.99.012306}. As illustrated in Fig.~\ref{fig:noisy_CQT_Xstate}, the protocol involves three parties: Alice, who acts as the sender, Bob, who is the receiver, and Charlie, who acts as the controller. Alice possesses an arbitrary single-qubit input state
\begin{equation}
|\psi_{\mathrm{in}}\rangle=\alpha|0\rangle+\beta|1\rangle,
\qquad |\alpha|^2+|\beta|^2=1,
\end{equation}
while Alice, Bob, and Charlie share a three-qubit entangled resource (For the ideal case without noise, the maximally entangled GHZ state is $\rho_{GHZ}=|GHZ\rangle \langle GHZ|$ with $|GHZ\rangle=1/\sqrt{2}(|000\rangle+|111\rangle)$) whose qubits are labeled $1$, $2$, and $3$, respectively. In the present study, this shared state is subjected to noise before being used as a resource for CQT. Charlie's role is to control the teleportation process by providing the additional classical information Bob needs to reconstruct the input state. Thus, unlike conventional two-party teleportation \cite{PhysRevLett.70.1895}, successful state transfer in CQT depends on the cooperation of the third party \cite{PhysRevA.58.4394,Barasinski2018,PhysRevLett.122.170501,PhysRevA.99.012306}. 

The CQT protocol proceeds in three main steps. First, Alice performs a Bell-state measurement on the unknown input qubit ($|\psi_{\mathrm{in}}\rangle$) and her qubit $1$ of the shared three-qubit GHZ resource through a noisy quantum channel. The Bell measurement projects the two qubits onto one of the four Bell states and produces two classical bits of information, denoted by $c_0$ and $c_1$. These measurement outcomes are communicated to Bob through a classical channel. The two bits determine the Pauli correction that Bob must apply to account for Alice's measurement outcome.

Next, Charlie performs the controller operation on his qubit $3$ of \textit{GHZ} state, which is shared through the noisy quantum channel. Specifically, he applies a basis rotation $R_{\theta}$ followed by a measurement in the computational ($Z$) basis. The measurement produces an additional classical bit, denoted by $c_2$, which is also classically communicated to Bob. This controller measurement is essential to the protocol's controlled nature, since Bob does not possess all the information required to reconstruct the input state unless Charlie cooperates. For the optimal CQT protocol and allowing the teleportation, the controller rotation is chosen as $\theta=\pi/4$ \cite{PhysRevLett.122.170501,PhysRevA.99.012306}.

Finally, Bob uses the classical information he receives from Alice and Charlie to apply a conditional local unitary operation to his qubit $2$ of \textit{GHZ}, which is again shared through the noisy quantum channel. The correction operation can be written as
\begin{equation}
U_{\mathrm{B}}=Z^{c_0}X^{c_1}Z^{c_2},
\end{equation}
where the exponents $c_0$, $c_1$, and $c_2$ are determined by the corresponding measurement outcomes. After applying this operation, Bob obtains the output state $\rho_{\mathrm{out}}$. In the absence of noise and for an ideal GHZ resource, the protocol allows Bob to recover the arbitrary input state with unit fidelity $\langle\Psi_{\rm in}|\rho_{\rm out}|\Psi_{\rm in}\rangle=1$ when Charlie allows. More generally, for the noisy three-qubit $X$-state considered in this work, the recovered state depends on the noisy resource's populations and coherences.

The schematic in Fig.~\ref{fig:noisy_CQT_Xstate} therefore represents the complete information flow of the CQT protocol: Alice supplies the unknown input state $|\psi_{\mathrm{in}}\rangle$ and performs the sender-side measurement, Charlie controls the transfer through his measurement and classical communication, and Bob applies the corresponding conditional operation to obtain $\rho_{\mathrm{out}}$.

\subsection{Proof-of-Concept Experimental proposal}
\begin{figure}[htbp]
    \includegraphics[width=\textwidth]{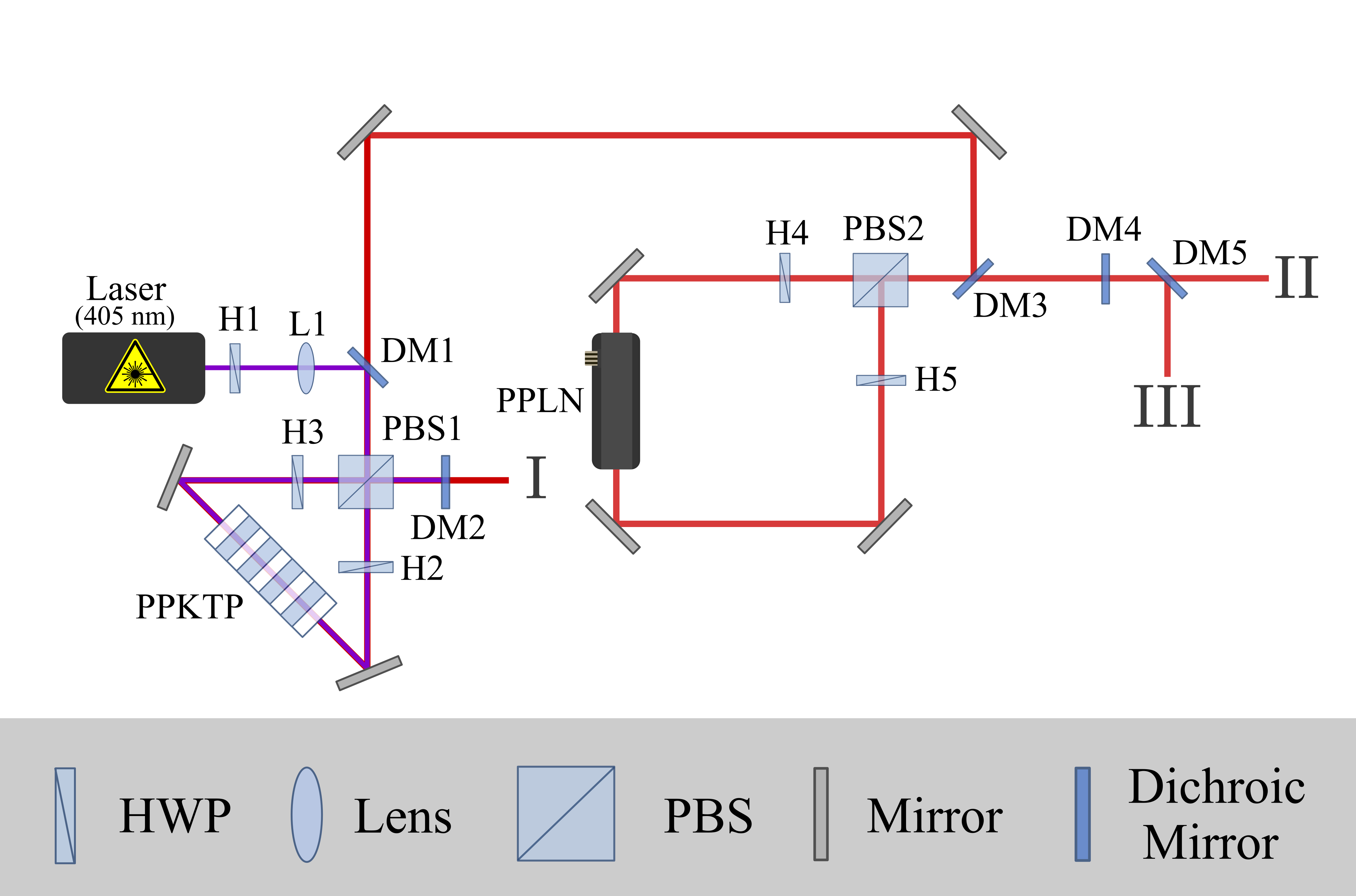}
    \caption{Schematic representation of an entangled photon source unit configured to generate a three-photon (I, II, and III) GHZ state.}
    \label{3GHZ_source}
\end{figure}

The physical realization of a three-qubit photonic channel is experimentally well motivated where the horizontal ($H$) and vertical ($V$) polarization states serve as the two logical basis states of each photonic qubit, i.e., $|H\rangle\equiv|0\rangle$ and $|V\rangle\equiv|1\rangle$. In particular, polarization-encoded GHZ states can be generated using cascaded spontaneous parametric down-conversion (C-SPDC) \cite{PhysRevA.105.063705}. A phase-stable implementation based on cascaded Sagnac interferometers has demonstrated three-photon polarization entanglement with fidelity exceeding $96\%$ relative to the ideal GHZ state $|GHZ\rangle=1/\sqrt{2}(|HHH\rangle+|VVV\rangle)$, while maintaining the state preparation without continuous active phase optimization. Such a three-qubit GHZ source in Fig.~\ref{3GHZ_source} provides a practical route for preparing the three-photon entangled resource assumed in our analysis, and its intrinsic phase stability \cite{PhysRevA.105.063705} is especially relevant when assessing the effects of decoherence and channel noise.

A grating-stabilized, temperature-controlled continuous-wave laser operating at approximately 405 nm provides the pump field. The pump polarization is controlled using a Half Wave Plate (HWP H1 set at 45 degrees) and subsequently focused by lens L1 to produce a state $1/\sqrt{2}(|H\rangle+|V\rangle)$ before being directed through the dichroic mirror (DM1) into the first Sagnac interferometer. The first interferometer consists of a Polarizing Beam Splitter (PBS1), H2 at 45 degrees, H3 at 0 degrees, steering mirrors, and a PPKTP crystal. This establishes the counter-propagating polarization paths, while the HWPs provide the required polarization transformations. Bidirectional excitation of the PPKTP crystal produces polarization-entangled photon pairs $1/\sqrt{2}(|HV\rangle+|VH\rangle)$ through type-II SPDC.

The generated photons are separated by PBS1, with one photon routed through DM2 to the I output channel, while the other is directed along the subsequent optical path toward a second Sagnac interferometer. The second interferometer comprises PBS2, H4 at 45 degrees, H5 at 0 degrees, steering mirrors, and a PPLN waveguide crystal. H4 and H5 control the polarization conditions required for bidirectional pumping, while PBS2 establishes the counter-propagating paths. The PPLN waveguide operates in the type-0 SPDC regime and generates a second photon pair. The generated photons are separated by DM3, DM4, and DM5, and routed to output channels II and III. Consequently, output channel I contains the photon generated in the first type-II SPDC process, whereas output channels II and III contain the two photons generated in the subsequent type-0 SPDC process, coincidence counting of which forms a three-photon GHZ state $\rho_{GHZ}$.

\begin{figure}[htbp]
    \includegraphics[width=\textwidth]{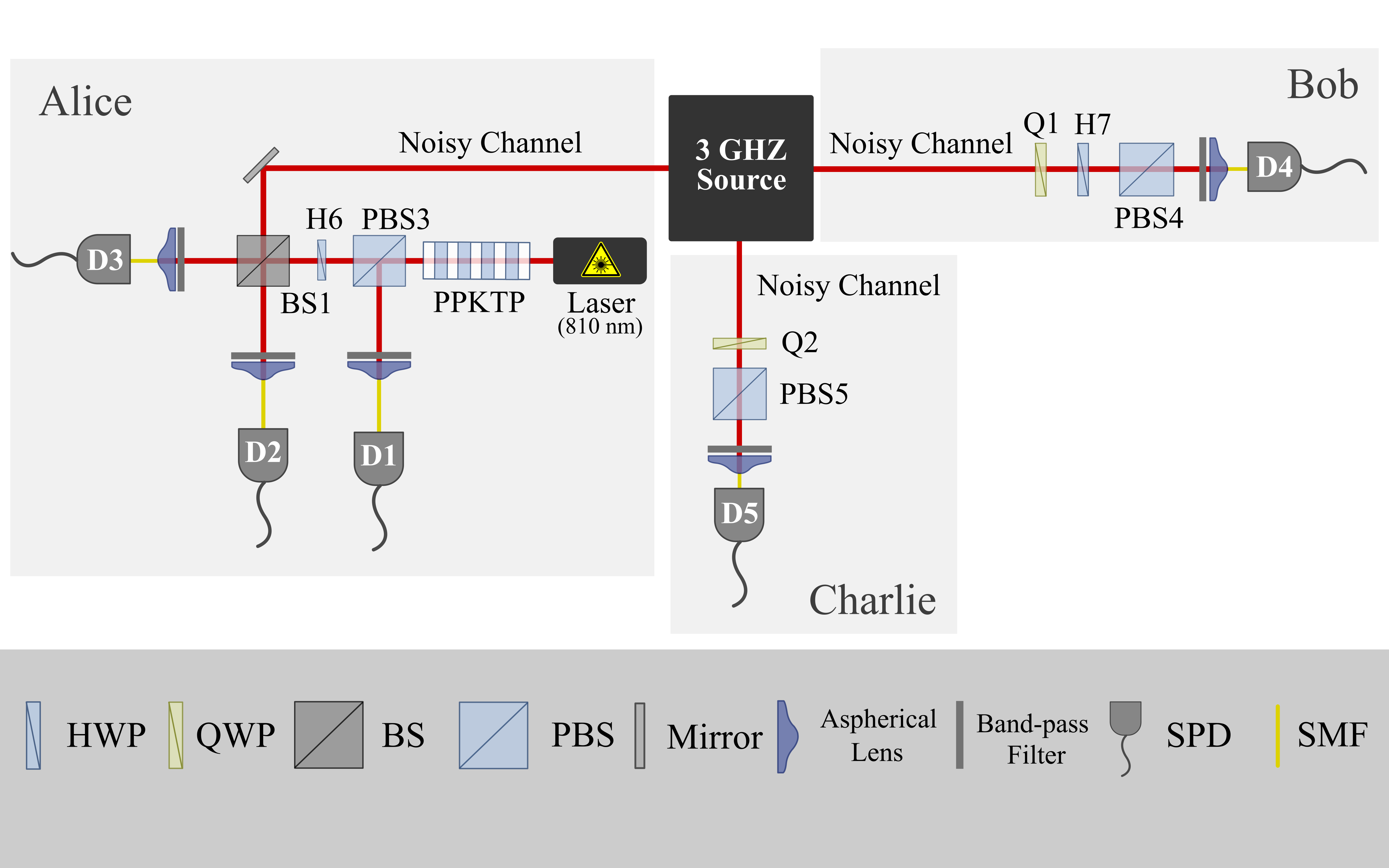}
    \caption{Schematic representation of a CQT arrangement employing a three-photon GHZ source (I, II, and III) through noisy-channel paths for Alice, Bob, and Charlie.}
    \label{CQT_setup}
\end{figure}

In the CQT configuration based on Ref. \cite{PhysRevLett.122.170501}        (Fig.~\ref{CQT_setup}), the incoming photon from the three-qubit GHZ source is directed to Alice through noisy channel, where it is combined with a photon $|\psi_{\mathrm{in}}\rangle=\alpha|H\rangle+\beta|V\rangle$ qubit to be teleported, with $\alpha=\text{cos}\theta$ and $\beta=\sqrt{1-\alpha^2}$. This photon is generated by the 810-nm laser and the PPKTP source. The generated photon pair is separated using PBS3, with H6 at $\theta$ degrees providing polarization-state control ($\alpha$ of $|\psi_{\mathrm{in}}\rangle$); the heralding photon is directed through PBS3 to the single-photon detector D1, while the other photon is combined with the incoming GHZ photon through noisy channel at the Beam Splitter BS1 for a partial Bell-state measurement. The Bell-state projection is realized through the interference of the two photons at BS1 and coincidence counting at D2 and D3. In parallel, the controller, Charlie, receives another photon of the three-qubit GHZ state through a separate noisy channel, where the angle of Q2 ($R_{\theta}$) determines if the teleportation protocol is allowed (QWP at 45°) or aborted (QWP at 0°) and is accordingly recorded by D5. The correlated detection outcomes at D2, D3, and D5 provide the measurement information required to implement and verify the controlled quantum teleportation protocol. Finally, another photon of the three-qubit GHZ state is transmitted through the noisy channel to Bob, where Q1, H7, and PBS4 perform Pauli correction ($U_{\mathrm{B}}$ from Eq. (2)) before detection by D4.

\section{Quantum Noise Channels}
The influence of noisy channels on the three-qubit resource ($\rho_{GHZ}$) is further described using the Kraus operator formalism. The evolution of the noisy state $\rho_X$ is given by
\begin{equation}
\rho_X=\sum_{i}^{n_i}\sum_{j}^{n_j}\sum _{k}^{n_k} E_{ijk}~\rho_{GHZ}~E_{ijk}^\dag~~;~~i,j,k=1,2,3 \cdots
\end{equation}
where $E_{ijk}$ are the three-qubit Kraus operators $E_{ijk}=\sum_{i,j,k=1}^{n_{i,j,k}}E_i(p)\otimes E_j(p)\otimes E_k(p)$ with single-qubit Kraus operators $E_i$ satisfying the completeness relation i.e.,  $\sum_i E_i^\dag E_i=\mathds{1}$, with $\mathds{1}$ denotes $2\times 2$ identity operator and \(n_{i,j,k}\) denote the number of Kraus operators required to represent the noise channel. 

\subsection{Bit-flip Noise}
The bit-flip channel changes the state of a qubit from $|0\rangle$ to $|1\rangle$ or from $|1\rangle$ to $|0\rangle$ with probability $p_i$. The corresponding Kraus operators are
\begin{equation}
E_1=\sqrt{1-\frac{p_i}{2}}\mathds{1},
\qquad
\qquad E_2=\sqrt{\frac{p_i}{2}} \sigma_x. 
\end{equation}
\subsection{Phase-flip Noise}
The phase-flip channel changes the phase of the state $|1\rangle$ to $-|1\rangle$ with probability $p_i$. Its Kraus operators are

\begin{equation}
E_1=\sqrt{1-p_i}\mathds{1},
\qquad
E_2=\sqrt{p_i}\sigma_z.
\end{equation}
This noise disrupts quantum coherence by suppressing the off-diagonal elements of the density matrix, while keeping the populations unchanged.
\subsection{Depolarising noise}
The depolarising channel replaces the qubit state with the maximally mixed state $\mathds{1}/2$ with probability $p_i$. Its Kraus operators are
\begin{equation}
 E_1=\sqrt{1-\frac{3p_i}{4}}\mathds{1},
\qquad
    E_2=\sqrt{\frac{p_i}{4}}\sigma_x,
\end{equation}
\begin{equation}
E_3=\sqrt{\frac{p_i}{4}}\sigma_y,
\qquad
E_4=\sqrt{\frac{p_i}{4}}\sigma_z, 
\end{equation} where $\sigma_x$, $\sigma_y$ and $\sigma_z$ are the Pauli matrices.
\subsection{Amplitude-damping Noise}
The amplitude-damping channel models energy dissipation from the excited state $|1\rangle$ to the ground state $|0\rangle$. The Kraus operators are
\begin{equation}
E_1=
\left(
\begin{array}{cc}
1 & 0\\
0 & \sqrt{1-p_i}
\end{array}
\right),
\qquad
E_2=
\left(
\begin{array}{cc}
0 & \sqrt{p_i}\\
0 & 0
\end{array}
\right),
\end{equation}
where the parameter $p_i$ represents the decay probability from the excited state to the ground state. This channel is widely used to describe spontaneous emission and other dissipative processes in quantum systems. 

For photonic systems, this dissipative model has a direct connection with
photon loss in optical-fiber transmission. In particular, photon loss in a
fiber can be described as a nonunitary decoherence process using a
Kraus-operator representation, and the resulting single-photon loss channel
has a direct analogy with the amplitude-damping channel
\cite{PhysRevA.105.062437}. According to the Beer--Lambert law, the
transmission probability of a photon through a fiber of length $L$ is
\begin{equation}
\eta(L)=10^{-\alpha L/10},
\end{equation}
where $\alpha$ is the fiber attenuation coefficient, conventionally
expressed in dB/km. The corresponding photon-loss probability is therefore
\begin{equation}
p_i(L)=1-\eta(L)
=1-10^{-\alpha L/10}.
\end{equation}
Thus, the amplitude-damping parameter $p_i$ can be given a direct physical
interpretation in terms of photon loss during fiber transmission, with the
effective noise strength determined by the fiber length and attenuation
coefficient \cite{PhysRevA.105.062437}.

With the application of these channels, the resulting
density matrices of noisy states (Appendix~A) have an X-shaped structure given as \begin{equation}
\rho_X =
\begin{pmatrix}
a_1 & 0 & 0 & 0 & 0 & 0 & 0 & z_1 \\
0 & a_2 & 0 & 0 & 0 & 0 & z_2 & 0 \\
0 & 0 & a_3 & 0 & 0 & z_3 & 0 & 0 \\
0 & 0 & 0 & a_4 & z_4 & 0 & 0 & 0 \\
0 & 0 & 0 & z_4^{*} & b_4 & 0 & 0 & 0 \\
0 & 0 & z_3^{*} & 0 & 0 & b_3 & 0 & 0 \\
0 & z_2^{*} & 0 & 0 & 0 & 0 & b_2 & 0 \\
z_1^{*} & 0 & 0 & 0 & 0 & 0 & 0 & b_1
\end{pmatrix}.
\label{eq:noisy_xstate}
 \end{equation}

\section{Quantum Correlations}

We assess the suitability of the noisy three-qubit GHZ channel for CQT by characterizing four complementary quantities: multipartite quantum discord, genuine multipartite concurrence (GMC), localizable entanglement (LE), and CQT fidelity. Together, these quantities characterize the nonclassical correlations (discord), GME (GMC), averaged conditional biseparable entanglement accessible through the controller (LE), and operational teleportation performance (CQT fidelity) of the noisy channel.

\subsection{Multipartite Quantum Discord}
\label{sec:discord}

We quantify multipartite nonclassical correlations using the multipartite quantum discord introduced in Ref.~\cite{Radhakrishnan2020}, as an extension of original bipartite discord \cite{PhysRevLett.88.017901,LHenderson_2001}. For the noisy tripartite state $\rho_X$, it is defined through conditional projective measurements as
\begin{equation}
D_{A;B;C}\left(\rho_X\right)=\min_{\Pi^{AB}}\left[-S_{BC|A}\left(\rho_X\right)+S_{B|\Pi^{A}}\left(\rho_X\right)+S_{C|\Pi^{AB}}\left(\rho_X\right)\right],
\end{equation}
where $S=-\text{Tr}\rho \text{log}\rho$ denotes the von Neumann entropy and $\Pi^{AB}$ denotes the conditional measurement in which the basis of the measurement on $B$ may depend on the outcome obtained from $A$. All the logarithms are base 2. The discord is a non-negative quantity and vanishes for classically correlated states.
\subsection{Genuine Multipartite Entanglement}
\label{sec:gme}

To quantify GME, we use the GMC for $n$-qubit $X$ states \cite{PhysRevA.86.062303}. Since the noisy states $\rho_X$ ($n=3$) retain the $X$-state structure, the GMC is given directly by
\begin{equation}
\mathrm{GMC}=2\max\left\{0,\left|z_i\right|-\omega_i\right\},\qquad i=1,\ldots,N,
\end{equation}
where $N=2^{n-1}$, $z_i$ are the anti-diagonal elements of the $X$-state, and
\begin{equation}
\omega_i=\sum_{j\neq i}^N \sqrt{a_j b_j},
\end{equation}
with $a_j$ and $b_j$ denoting the diagonal elements of the state. GMC is generalized to both pure and mixed states and satisfies $\mathrm{GMC}>0$ for genuinely tripartite entangled states (GHZ and W class), whereas $\mathrm{GMC}=0$ indicates the absence of genuine tripartite entanglement (biseparable and separable class) \cite{7cfj-8klm}. For a two-qubit state ($n=2$), this GMC reduces to Wootters concurrence \cite{PhysRevLett.78.5022,PhysRevLett.80.2245}.

\subsection{Localizable Entanglement}
\label{Sec_LocEnt}

The loss of GMC does not by itself determine whether the noisy state $\rho_X$ remains useful for CQT \cite{Barasinski2018,PhysRevA.99.012306}. We therefore quantify the entanglement that can be localized between the sender $S$ and receiver $R$ by optimal measurements on the controller $C$ which is expressed as \cite{Barasinski2018},
\begin{equation}
\mathcal{C}_L\left(\rho_X\right) = \max_{U_C}\left[\sum\limits_{t=0}^{1}\langle t| U_C \rho_C U_C^{\dagger}|t\rangle\mathrm{GMC}\left(\rho_{SR}^{t}\right)\right] \,,
\label{eq:locconc}
\end{equation}
where optimization is performed over all $2\times2$ unitary matrices $U_C$ with reduced state $\rho_C=\text{Tr}_{SR}\left(\rho_X\right)$, and the state of subsystem $SR$ is
\begin{equation}
    \rho_{SR}^{t} = \dfrac{\text{Tr}_{C}\left[U_C^{\dagger}|t\rangle \langle t|U_C \otimes \mathds{1}_4 \left(\rho_X\right) U_C^{\dagger}|t\rangle \langle t|U_C \otimes \mathds{1}_4\right]}{\langle t|U_C\rho_C U_C^{\dagger}|t\rangle} \,.
\end{equation}
$\rho_{SR}^{t}$ is obtained after admissible local measurement $U_C$ on the controller qubit $C$ with outcome $t$ and ${\langle t|U\rho_C U_C^{\dagger}|t\rangle}$ is the probability of obtaining the outcome. $\mathcal{C}_{L}$ measures the maximum averaged conditional biseparable entanglement that can be made available to sender and receiver through the controller's measurement. Importantly, $\mathrm{GMC}=0$ does not imply $\mathcal{C}_{L}=0$ because biseparable mixed states can retain nonzero $\mathcal{C}_{L}$ and can therefore remain useful for CQT, including cases in which the teleportation fidelity exceeds the classical limit \cite{Barasinski2018,7cfj-8klm}. 

\subsection{Controlled Quantum Teleportation Fidelity}
\label{sec:fidelity-measures}

Teleportation fidelity is optimal for $\theta=\pi/4$ in the controller's $R_\theta$ \cite{PhysRevA.58.4394} and is given as \cite{Barasinski2018,PhysRevA.99.012306,PhysRevLett.122.170501}
\begin{equation}
\mathcal{F}\left(\rho_X\right)
= \max_{i=1\ldots4}F_{\rm CQT}^{(i)}\left(\rho_X\right),\\
\label{Eq_15}
\end{equation}

\begin{equation}
\begin{aligned}
  F_{\rm CQT}^{(1)}\left(\rho_X\right) &= \frac{3 + |\Delta_1| + 4(|z_1| + |z_4|)}{6}, F_{\rm CQT}^{(2)}\left(\rho_X\right) = \frac{3 + |\Delta_1| + 4(|z_2| + |z_3|)}{6},\\
  F_{\rm CQT}^{(3)}\left(\rho_X\right) &= \frac{3 + \sqrt{|\Delta_2|^2 + 16(|z_1| + |z_4|)^2}}{6}, F_{\rm CQT}^{(4)}\left(\rho_X\right) = \frac{3 + \sqrt{|\Delta_2|^2 + 16(|z_2| + |z_3|)^2}}{6},
\end{aligned}
\label{Eq_16}
\end{equation}
where $\Delta_1 = a_1 - a_2 - a_3 + a_4 + b_1 - b_2 - b_3 + b_4$, and
$\Delta_2 = a_1 - a_2 + a_3 - a_4 - b_1 + b_2 - b_3 + b_4$.
Here, $a_i$, $b_i$, and $z_i$ are defined in Eq.~(\ref{eq:noisy_xstate}). The resulting $F_{CQT}$ is then compared with the classical teleportation threshold $F_{\mathrm{cl}}=2/3$ to determine whether the noisy channel retains a quantum advantage.

\section{Results and Discussion}

In what follows, we summarize the results by examining the four quantities: multipartite discord, GMC, LE and CQT fidelity. For each quantity, we investigate the effects of different noisy channels on the GHZ resource with equal damping acting on all the qubits (sender, receiver and controller, $p_i=p$ for $i=1,2,3$ in Eqs. (4)-(8)) and a pair of qubits (receiver and controller, $p_i=p$ for $i=2,3$ in Eqs. (4)-(8), $p_1=0$). We then compare the behavior of these measures to elucidate the interplay between quantum correlations. 

\begin{figure}[htbp]
    \centering

    \begin{subfigure}{0.495\textwidth}
        \centering
        \includegraphics[width=\linewidth]{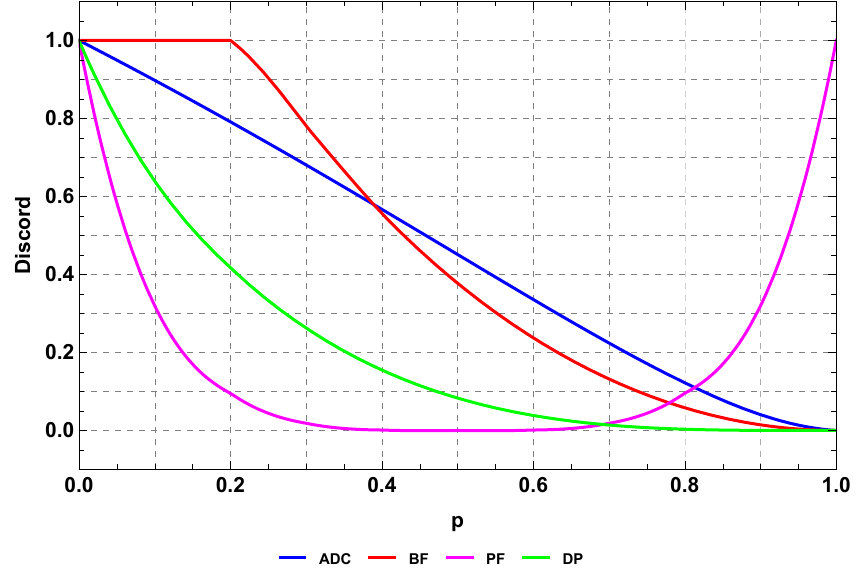}
        \caption{Noise on receiver and controller qubits}
     
    \end{subfigure}
 \begin{subfigure}{0.495\textwidth}
        \centering
        \includegraphics[width=\linewidth]{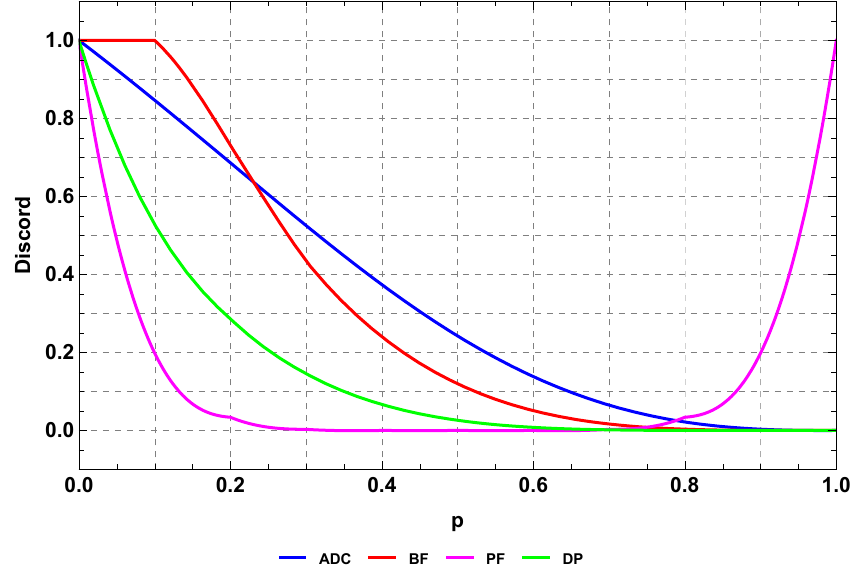}
        \caption{Noise on sender, receiver and controller qubits}
     
    \end{subfigure}
\caption{Quantum discord as a function of the noise strength \(p\) under amplitude-damping (ADC), bit-flip (BF), phase-flip (PF) and  depolarizing (DP) noise acting on the qubits.}
\end{figure}
Fig.~4 shows how the multipartite discord of the noisy GHZ state varies with the noise strength $p$ for the four different noise channels. Amplitude damping and bit-flip yield the slowest decay of discord and only at large noise parameters $p$ does discord approach zero. The bit-flip noise curves exhibit a freezing phenomenon, with the discord remaining at its maximum value up to a finite noise strength of $p \approx 0.20$ when noise acts on the receiver and controller qubits [Fig. 4(a)] and $p \approx 0.10$ when all qubits are subjected to noise [Fig. 4(b)]. Beyond these thresholds, the discord decreases monotonically, approaching zero as $p \to 1$. Furthermore, the discord is most vulnerable to phase flip and depolarizing noise. For phase-flip noise, the discord falls quickly from maximal value to zero as $p$ increases to 0.50 and shows weak quantum correlations close to zero over an intermediate range ($p\approx0.40$–0.60 for the case of noise on receiver \& controller qubits in Fig. 4 (a); $p\approx0.35$–0.65 for the case of noise on all qubits in Fig. 4 (b)). For the case of depolarizing channel with the two noisy qubits in Fig. 4 (a), discord approaches zero by $p\approx0.86$; with all three noisy qubits in Fig. 4 (b), discord approaches zero at $p\approx0.74$. Notably, the freezing interval becomes shorter when noise acts on all three qubits, indicating that the robustness of multipartite discord depends on the number of qubits exposed to environmental noise. The existence of a finite freezing regime demonstrates that nonclassical correlations can remain insensitive to increasing bit-flip noise over a limited parameter range. Such noise-resilient correlations may be relevant to quantum information processing tasks \cite{freezing1,freezing2,freezing3}, although their persistence alone does not guarantee an operational quantum advantage.

\begin{figure}[H]
    \centering

    \begin{subfigure}{0.495\textwidth}
        \centering
        \includegraphics[width=\linewidth]{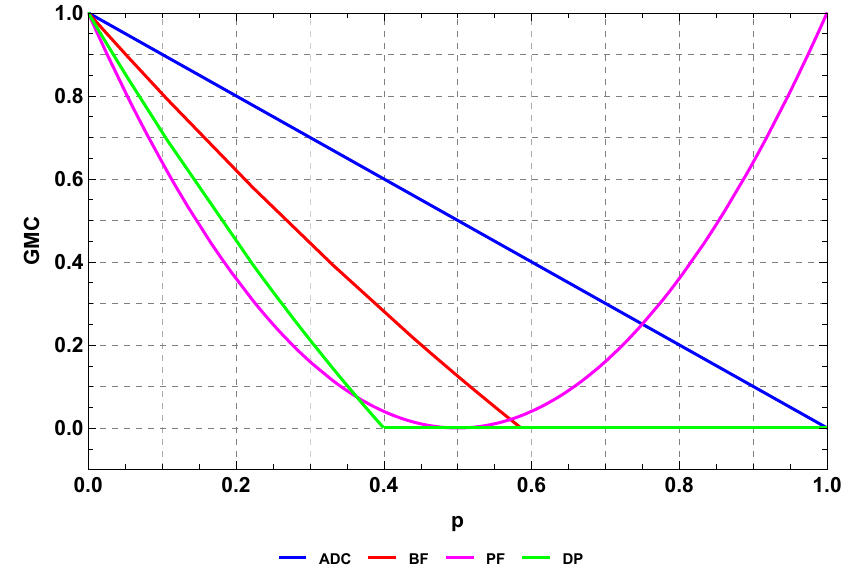}
        \caption{Noise on receiver and controller qubits}
     
    \end{subfigure}
 \begin{subfigure}{0.495\textwidth}
        \centering
        \includegraphics[width=\linewidth]{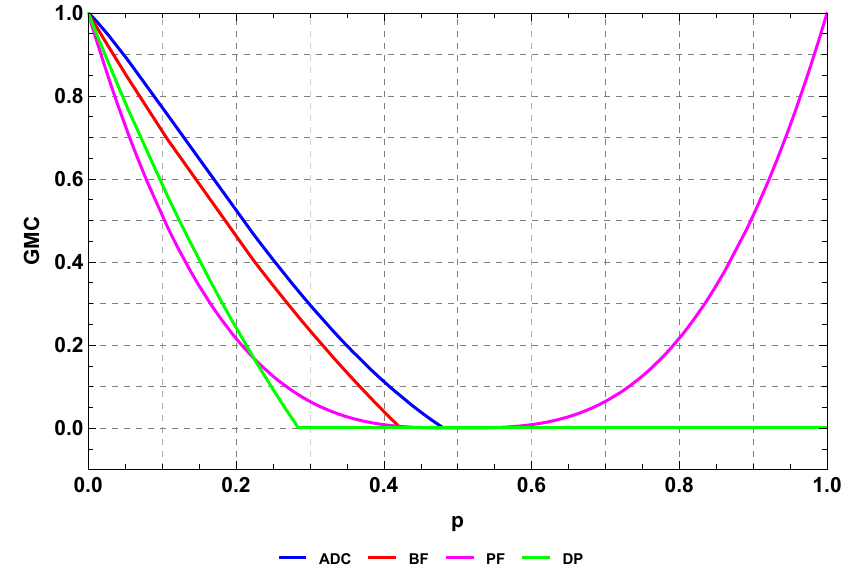}
        \caption{Noise on sender, receiver and controller qubits }
  
    \end{subfigure}
      \caption{Genuine multipartite concurrence (GMC) as a function of the noise strength \(p\) under amplitude-damping (ADC), bit-flip (BF), phase-flip (PF) and  depolarizing (DP) noise acting on the qubits.}
\end{figure}
We next examine the GMC of the noisy GHZ state in Fig.~5. Unlike the gradual decay behavior observed for multipartite discord in Fig.~4, GMC exhibits sudden death behavior reaching zero at a finite value of $p$ for several of the noise models considered. This GMC behavior was previously observed for the bit-flip and amplitude damping noise on  GHZ state \cite{Sk2022}. Similar to discord, amplitude-damping and bit-flip noise preserve GMC over a broader parameter range than depolarizing noise.  However, bit-flip noise leads to the earlier sudden death of GMC when compared with amplitude damping behavior. With bit-flip noise on receiver and controller qubits, GMC falls to zero around $p\approx0.59$ in Fig. 5 (a) and with all three qubits noisy, it vanishes earlier, around $p\approx0.43$ in Fig. 5 (b). Interestingly, multipartite discord remains nonzero beyond the sudden death of GMC. This demonstrates that the disappearance of GME does not necessarily imply the complete loss of nonclassical correlations in the noisy GHZ resource. Consequently, multipartite discord provides complementary information about the quantum-correlation structure that cannot be captured by GMC alone. Introducing additional noise on the sender's qubit accelerates the degradation of GME, resulting in an earlier disappearance of GMC. For the amplitude damping noise on two qubits case, GMC decreases to zero only in the limit $p\to1$ in Fig. 5 (a) and with noise on all three qubit GMC undergoes sudden death at a finite $p\approx0.48$ in Fig. 5 (b). For phase-flip noise GMC reaches zero at $p=0.5$ in Fig.s 5 (a), (b).  GMC is extremely fragile to depolarizing noise. With noise on two qubits GMC reaches zero at $p\approx0.40$ in Fig. 5 (a) and with noise on all three qubits it reaches zero even earlier at $p\approx0.27$ in Fig. 5 (b).

\begin{figure}[htbp]
    \centering

    \begin{subfigure}{0.495\textwidth}
        \centering        \includegraphics[width=\linewidth]{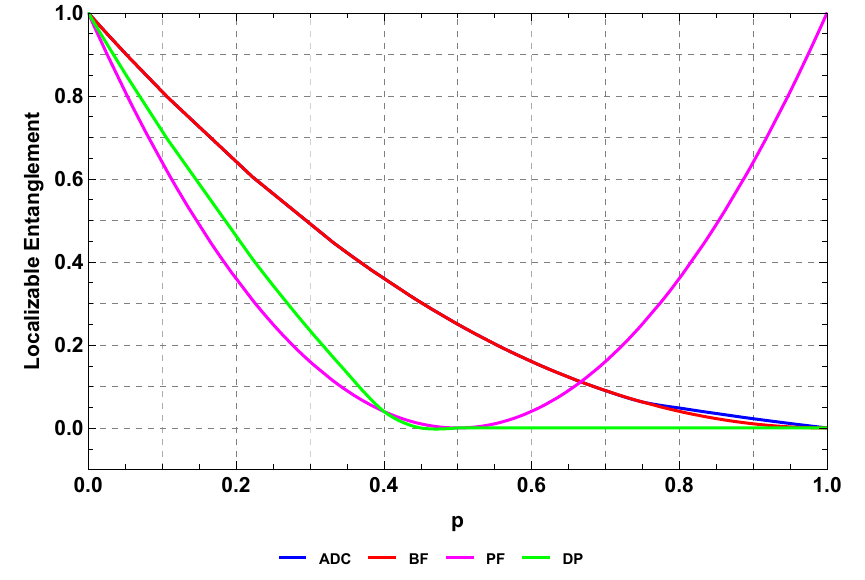}
        \caption{Noise on receiver and controller qubits}
     
    \end{subfigure}
 \begin{subfigure}{0.495\textwidth}
        \centering
        \includegraphics[width=\linewidth]{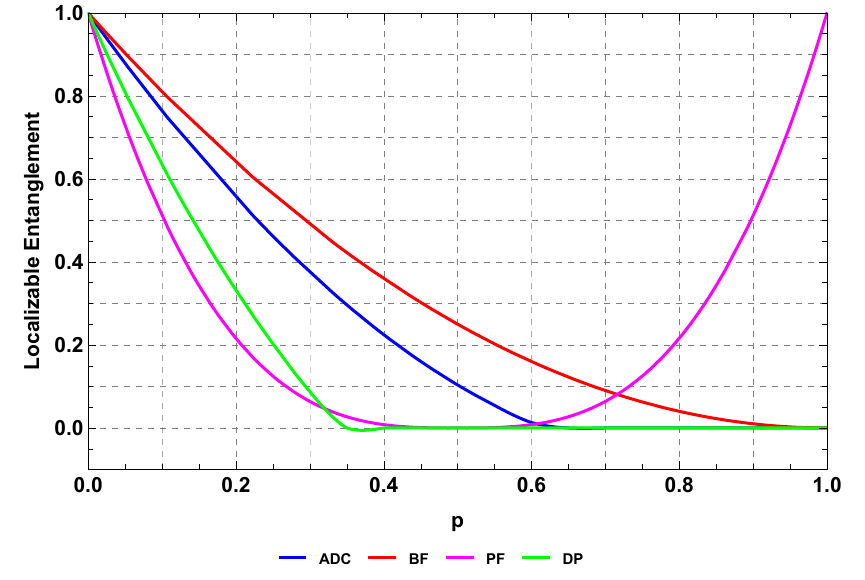}
        \caption{Noise on sender, receiver and controller qubits}
    \end{subfigure}
    \caption{Localizable entanglement as a function of the noise strength \(p\) under amplitude-damping (ADC), bit-flip (BF), phase-flip (PF) and  depolarizing (DP) noise acting on the qubits.}
\end{figure}

 Fig. 6 plots LE vs.~$p$ behaviour with the case of noise on receiver and controller qubits in Fig. 6 (a) and Fig. 6 (b) shows the case of noise on sender, receiver and controller qubits. LE is particularly robust against bit-flip noise and in both cases, LE decreases slowly and remains non zero up to $p=1$. Thus, LE is not completely destroyed by bit-flip noise similar to multipartite discord, even though the GMC undergoes sudden death at finite values of $p$, as shown in Fig.~5. For amplitude damping noise on two qubits case in Fig. 6 (a), LE shows asymptotic decay behavior reaching zero only at $p=1$. For noise on all three qubits case in Fig. 6 (b), LE reaches zero at $p\approx0.65$ which is later than GMC death in Fig. 5 (b). Similarly, for depolarizing noise on two (three) qubits case in Fig. 6 (a) (Fig. 6 (b)), LE reaches zero at $p\approx0.45$ ($p\approx0.35$) which is later than GMC death in Fig. 5 (a) (Fig. 5 (b)).This clearly indicates that LE can survive in mixed noisy states (biseparable class) even after the loss of genuine tripartite entanglement (GHZ and W class) \cite{Barasinski2018,7cfj-8klm}. In case of phase-flip noise, LE shows similar behavior as GMC, unlike other noises.

\begin{figure}[htbp]
    \centering

    \begin{subfigure}{0.495\textwidth}
        \centering
        \includegraphics[width=\linewidth]{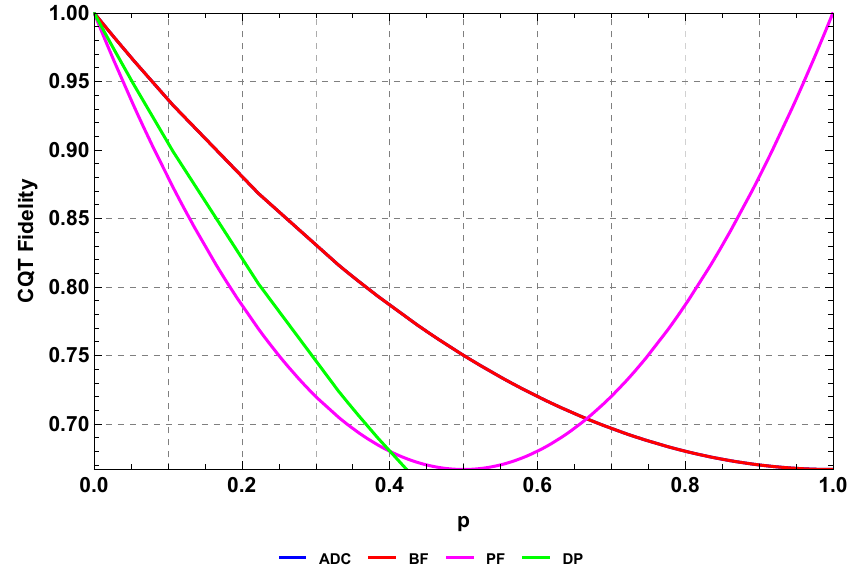}
        \caption{Noise on receiver and controller qubits}
     
    \end{subfigure}
 \begin{subfigure}{0.495\textwidth}
        \centering
        \includegraphics[width=\linewidth]{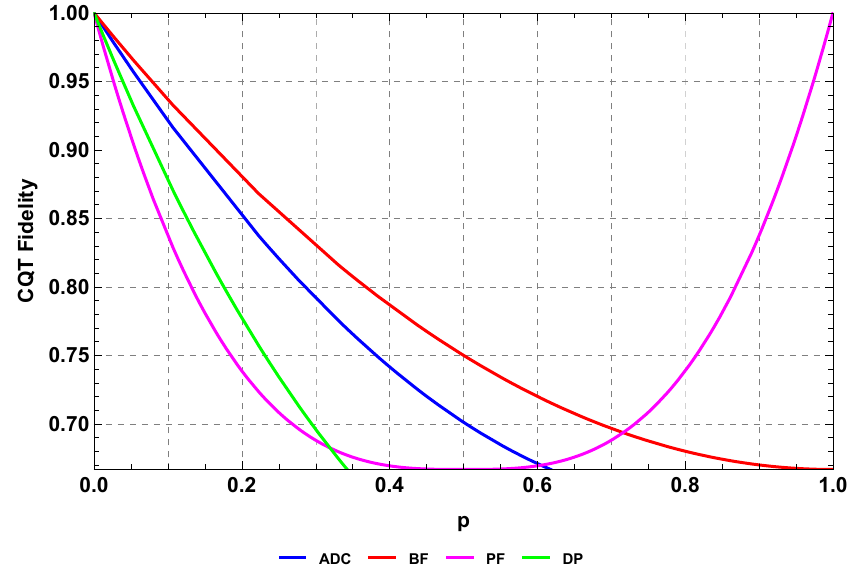}
        \caption{Noise on sender, receiver and controller qubits}
     
    \end{subfigure}
    \caption{Controlled quantum teleportation (CQT) Fidelity as a function of the noise strength \(p\) under amplitude-damping (ADC), bit-flip (BF), phase-flip (PF) and  depolarizing (DP) noise acting on the qubits.}
\end{figure}

Finally, Fig. 7 shows the CQT fidelity $\mathcal{F}$ vs. noise strength $p$, for teleporting an arbitrary single qubit utilizing the noisy GHZ state as a resource. The plots are shown above the classical benchmark $F_{\rm cl}=2/3$ with the case of noise on receiver and controller qubits in Fig. 7 (a) and Fig. 7 (b) shows the case of noise on sender, receiver and controller qubits. The four noisy states show matching behavior with LE in Fig. 6 and amplitude-damping shows the most robustness. This reinforces that GME is a sufficient resource but not a necessary one for CQT advantage. On the other hand, the maximum averaged conditional biseparable entanglement after Charlie’s (controller) measurement as captured by LE is a more relevant measure of the CQT resource.

A comparison of Fig.s 4–7 reveals that the robustness of the noisy GHZ resource depends not only on the type and strength of environmental noise but also on the qubits exposed to noise and the quantum-correlation measure considered. In particular, multipartite discord can survive beyond the sudden death of genuine multipartite concurrence, demonstrating the persistence of nonclassical correlations in the absence of genuine tripartite entanglement. Similarly, localizable entanglement may remain nonzero after the disappearance of genuine multipartite entanglement, indicating that biseparable entanglement can still be established between the sender and receiver through the controller's measurement. However, the persistence of nonclassical correlations does not necessarily guarantee an operational quantum advantage, which must be assessed independently through the controlled teleportation fidelity. Thus, multipartite discord provides a complementary characterization of the noisy resource, whereas localizable entanglement is more directly connected to its usefulness for controlled quantum teleportation. Including noise on the sender's qubit generally accelerates the degradation of controlled teleportation fidelity when compared to the case of noise only on receiver and controller qubits (assuming the GHZ resource was shared to Bob and Charlie by the sender Alice with negligible noise on her qubit). These findings highlight the distinct roles of quantum correlations and operational resources in determining the resilience of GHZ-based quantum communication protocols.

\section{Conclusions}

We have investigated the robustness of three-qubit GHZ-based controlled quantum teleportation under bit-flip, phase-flip, depolarizing, and amplitude-damping noise by examining multipartite quantum discord, genuine multipartite concurrence (GMC), localizable entanglement (LE), and CQT fidelity. Our results reveal that the degradation of quantum correlations and teleportation performance depends on both the noise channel and the qubits exposed to noise. Notably, multipartite discord exhibits freezing under bit-flip noise over a finite parameter regime. Furthermore, discord persists beyond the sudden death of genuine multipartite entanglement, demonstrating the survival of nonclassical correlations even in the absence of genuine tripartite entanglement. Importantly, the disappearance of GMC does not necessarily eliminate the quantum advantage in CQT, while LE provides a more direct characterization of the entanglement accessible through the controller's measurement and more closely follows the CQT fidelity \cite{Barasinski2018,7cfj-8klm}. These findings establish the complementary roles of multipartite discord, multipartite entanglement, and localizable entanglement \cite{PhysRevResearch.4.023035} in characterizing noisy multipartite quantum resources. 

The proposed photonic implementation offers a potential route for experimental investigation, while future extensions could explore temporally correlated noise and memory effects \cite{Sk2022} in multipartite quantum communication.

\bibliography{sample}

\section*{Appendix}
\section*{Matrix elements of the noisy GHZ states}
For all four channels the noisy resource $\rho_X$ keeps the X-shaped structure of Eq.~(11).
Its elements are fixed by the three noise parameters $p_1,p_2,p_3$ (qubits 1, 2, 3 denote Alice, Bob, Charlie).
\subsection*{Bit Flip noise}

\[
\begin{aligned}
a_1=b_1=z_1=z_1^{*}&=\frac{1}{8}
\left((p_2-2)(p_3-2)+p_1(p_2+p_3-2)\right),\\
a_2=b_2=z_2=z_2^{*}&=\frac{1}{8}
\left(p_1(p_2-p_3)-(p_2-2)p_3\right),\\
a_3=b_3=z_3=z_3^{*}&=\frac{1}{8}
\left(p_1p_3-p_2(p_1+p_3-2)\right),\\
a_4=b_4=z_4=z_4^{*}&=\frac{1}{8}
\left(p_2p_3-p_1(p_2+p_3-2)\right).
\end{aligned}
\]
\subsection*{Phase Flip noise}

\[
\begin{aligned}
a_1=b_1&=\frac{1}{2},\\
z_1=z_1^{*}&=\frac{1}{2}(1-2p_1)(1-2p_2)(1-2p_3),\\
\end{aligned}
\]
where the remaining elements are zero.

\subsection*{Amplitude Damping noise}
\[
\begin{aligned}
a_1&=\frac{1}{2}\left(1+p_1p_2p_3\right)\\
a_2&=\frac{1}{2}p_1p_2(1-p_3)\\
a_3&=\frac{1}{2}p_1(1-p_2)p_3\\
a_4&=\frac{1}{2}p_1(1-p_2)(1-p_3)\\
b_4&=\frac{1}{2}(1-p_1)p_2p_3\\
b_3&=\frac{1}{2}(1-p_1)p_2(1-p_3)\\
b_2&=\frac{1}{2}(1-p_1)(1-p_2)p_3\\
b_1&=\frac{1}{2}(1-p_1)(1-p_2)(1-p_3)\\
z_1&=z_2=z_1^{*}=z_2^{*}=\frac{1}{2}\sqrt{(1-p_1)(1-p_2)(1-p_3)}\\
\end{aligned}
\]

where the remaining off-diagonal elements vanish.

\subsection*{Depolarising noise}
\[
\begin{aligned}
a_1 &= b_1=\frac{1}{8}
\left(4-2p_1-2p_2-2p_3+p_1p_2+p_1p_3+p_2p_3\right),\\
a_2 &=b_2= \frac{1}{8}
\left(p_1p_2-p_1p_3-p_2p_3+2p_3\right),\\
a_3 &= b_3=\frac{1}{8}
\left(p_1p_3-p_1p_2-p_2p_3+2p_2\right),\\
a_4 &=b_4= \frac{1}{8}
\left(p_1p_3-p_1p_2+p_2p_3+2p_1\right),\\
z_1 &= z_8 = z_1^{*} = z_8^{*}
= \frac{1}{2}(1-p_1)(1-p_2)(1-p_3).
\end{aligned}
\]

where the remaining off-diagonal elements are zero.
\end{document}